\documentclass{seg}
\usepackage{microtype}
\usepackage{multirow}

\title{Parameter-Efficient Adaptation of Pre-Trained Vision Foundation Models for Active and Passive Seismic Data Denoising}

\author{%
  Jiahua Zhao$^{1}$\thanks{Corresponding author: j.zhao@cyi.ac.cy} \and
  Umair bin Waheed$^{2}$ \and
  Jing Sun$^{3}$ \and
  Yang Cui$^{2}$ \and
  Nikos Savva$^{1,4}$ \and
  Eric Verschuur$^{5}$
}
\date{30 April 2026}

\begin{document}

\maketitle

\begin{center}
  \footnotesize
  $^{1}$Computation-based Science and Technology Research Center, The Cyprus Institute, Nicosia, Cyprus\\
  $^{2}$Department of Geosciences, College of Petroleum Engineering and Geosciences, King Fahd University of Petroleum \& Minerals, Dhahran, Saudi Arabia\\
  $^{3}$Department of Intelligent Systems, Faculty of Electrical Engineering, Mathematics and Computer Science, Delft University of Technology, Delft, Netherlands\\
  $^{4}$Department of Mathematics and Statistics, Faculty of Pure and Applied Sciences, University of Cyprus, Nicosia, Cyprus\\
  $^{5}$Department of Geoscience and Engineering, Faculty of Civil Engineering and Geosciences, Delft University of Technology, Delft, Netherlands
\end{center}

\begin{abstract}
  The increasing demand for high-resolution subsurface imaging and continuous Earth monitoring has led to an exponential growth in both active and passive seismic data, driven by dense geophone deployments, distributed acoustic sensing (DAS) arrays, and large-scale 2D and 3D surveys. This rapid data expansion exacerbates the challenge of suppressing complex noise while preserving signal fidelity. Traditional supervised deep learning methods are typically designed for task-specific settings, requiring large paired datasets and often suffering from domain shift when applied to new acquisition conditions. Foundation models offer a promising pathway for learning transferable representations that can mitigate these limitations. However, pre-training seismic foundation models from scratch remains highly challenging, as it requires massive domain-specific datasets and extensive computational resources. Therefore, we propose an efficient framework that repurposes general-purpose Vision Foundation Models (VFMs) for geophysical tasks using Parameter-Efficient Fine-Tuning. Specifically, our architecture employs a pre-trained VFM, namely a DINOv3 encoder, adapted via Low-Rank Adaptation (LoRA), enabling effective feature adaptation with minimal additional parameters. Moreover, to improve robustness under unseen field conditions in the absence of a ground truth, we introduce a kurtosis-guided, unsupervised test-time adaptation module that updates only the LoRA parameters during inference. This mechanism allows the model to self-calibrate to site-specific noise characteristics by identifying information-rich regions based on kurtosis and performing self-training without requiring labeled data. Experiments on exploration field seismic images from public datasets and DAS vertical seismic profiling data from the Utah FORGE site show that the proposed framework matches or exceeds the performance of domain-specific models. Additional evaluation on unseen cross-site datasets from a land survey in China and the Groß Sch\"onebeck geothermal site in Germany further demonstrates the strong generalization capability of the proposed framework and its effectiveness in signal-noise separation. The results highlight the potential of adapting pre-trained VFMs for data-intensive challenges in exploration seismology.
\end{abstract}

  \section{Introduction}

  Seismic monitoring form the backbone of subsurface imaging and are essential for applications such as energy resource exploration, carbon storage, and hazard assessment \cite[]{hudson2024nodearrays,oukili2024ccsseismic}. These applications encompass both active seismic surveys, in which controlled sources are used to illuminate the subsurface, and passive seismic monitoring, which relies on naturally occurring or ambient signals for continuous observation. However, the fidelity of seismic data, spanning conventional geophone arrays to modern distributed acoustic sensing (DAS) systems \cite[]{taha2025dasreview}, is consistently affected by complex noise environments. Coherent noise such as ground roll and multiples, together with erratic disturbances, including coupling noise in DAS, can degrade the signal-to-noise ratio and obscure critical geological features \cite[]{rashid2025das}. 

  Over the past decades, numerous geophysical processing methods have been developed to address these challenges. Traditional processing workflows typically rely on sequences of linear and non-linear filters, such as bandpass \cite[]{stein1983continuously}, $f-k$ \cite[]{gulunay1986fxdecon}, wavelet transform \cite[]{deighan1997ground}, and Radon transform \cite[]{russell1990noise1,russell1990noise2} methods. However, these approaches often struggle to separate signal from noise when their spectral or kinematic properties overlap. In addition, these conventional approaches are labor-intensive, requiring expert-driven parameter tuning for each dataset, which limits their scalability for the massive data volumes generated by modern continuous monitoring systems.

  In recent years, deep learning, particularly convolutional neural network, has created new opportunities for seismic processing \cite[]{yu2019deep,zhu2019phasenet,zhang2019unsupervised}. Data-driven models based on modern convolutional neural network architectures, such as the U-Net architecture \cite[]{ronneberger2015u} and its variants, have demonstrated a remarkable ability in learning complex non-linear mappings, among which is recovering clean seismic signals from noise-contaminated observations in denoising tasks \cite[]{zhu2019seismic,zhong2022seismic,wu2023attenuating,wang2025quadratic}. 
  
  However, most existing approaches rely on supervised learning under task-specific settings, assuming similar statistical properties between training and deployment data, which is rarely satisfied in practical geophysical applications. Consequently, supervised models are highly sensitive to domain shift \cite[]{hou2021machine}, marked by differences in the statistical distribution between the training data (source domain) and the test data (target domain), and usually exhibit significant performance degradation when applied to previously unseen field data. 
  
  In geophysical applications, such domain shifts are common and can be pronounced. Many deep learning models are trained on synthetic datasets generated through physics-based modeling because field data rarely provide paired clean references (ground truth) \cite[]{merrifield2022synthetic}. While synthetic data offer controlled conditions and valuable physical consistency, they may not fully capture the complexity of real-world noise processes \cite[]{cui2024ground}, such as site-dependent coupling noise and scattering effects in heterogeneous overburden structures. As a result, discrepancies between synthetic and field data distributions can limit model generalization when applied to real observations.

  In addition, substantial variability exists across different geological sites. A model trained on data from one geological environment often performs poorly when applied to another \cite[]{maguire2024generalization}. Generally speaking, these performance degradations may arise from differences in source wavelet frequency content, local attenuation behavior (geological differences), and site-specific noise characteristics. When supervised models are exposed to such domain shifts, their effectiveness deteriorates, frequently leading to the suppression of valid seismic signals (signal leakage) or the generation of artificial features (hallucinated artifacts) \cite[]{cai2025seismic}.  Although transfer learning and fine-tuning are widely adopted to mitigate these issues \cite[]{birnie2022transfer}, they typically require labeled data from the target domain. Such labeled datasets are rarely available in practical seismic processing scenarios, which limits the applicability of conventional supervised adaptation strategies in real-world deployments.

  In the broader field of artificial intelligence, foundation models have emerged as a promising pathway for improving cross-domain generalization. Large-scale models such as GPT-4 \cite[]{achiam2023gpt} and the Segment Anything Model \cite[]{kirillov2023segment} are pre-trained on billion-scale datasets using self-supervised learning, enabling them to learn universal feature representations that generalize across diverse downstream tasks. Inspired by this paradigm, the geophysical community has begun to develop foundation models for seismic applications. Recent work, such as the Seismic Foundation Model (SFM) by \cite{sheng2025seismic}, pre-trains a Vision Transformer (ViT) \cite[]{dosovitskiy2021vit} on millions of unlabeled seismic patches to learn domain-specific features. While promising, pre-training an SFM from scratch requires massive computational resources, which can limit accessibility. 
  
  Instead of building expensive domain-specific models from scratch, an alternative strategy is to leverage existing general-purpose Vision Foundation Models (VFMs), such as DINOv3 \cite[]{simeoni2025dinov3} and Swin Transformer V2 \cite[]{liu2022swin}, which have been pre-trained on millons to billions of natural images (e.g., ImageNet-1k \cite[]{deng2009imagenet}, LVD-1689M \cite[]{simeoni2025dinov3}). Although seismic data differ from natural images in physical origin, they share fundamental visual primitives, including edges, textures, and multi-scale structures \cite[]{fuchs2025foundation}. Prior studies suggest that the feature representations learned by VFMs can exhibit cross-domain transferability and may be adapted to seismic data \cite[]{guo2025cross}. 
  
  To enable efficient adaptation, we employ Parameter-Efficient Fine-Tuning (PEFT), specifically Low-Rank Adaptation (LoRA). LoRA keeps the large pre-trained backbone parameters frozen and introduces small, trainable low-rank matrices \cite[]{hu2022lora}. This approach reduces the number of trainable parameters by orders of magnitude, enabling adaptation with limited computational resources while preventing catastrophic forgetting of the pre-trained knowledge.

  However, even with a powerful VFM backbone, the domain shift between the training data and a previously unseen target dataset remains a significant challenge. To address this issue without requiring ground-truth labels, we introduce a Kurtosis-guided test-time adaptation (TTA) module. Seismic data are typically sparse and exhibit non-Gaussian statistical characteristics, where high kurtosis values are associated with reflection events \cite[]{cui2025unsupervised}, whereas random noise tends to follow a more Gaussian distribution. Our TTA module exploits this statistical property to guide input selection during inference. Specifically, it identifies information-rich, high-kurtosis patches and uses them to update the LoRA trainable parameters in an unsupervised manner before processing the full volume. This mechanism allows the model to self-calibrate to the specific noise characteristics of the new dataset, thereby mitigating domain discrepancies during deployment.

  This study aims to develop, validate, and benchmark an efficient framework for seismic denoising that improves robustness across datasets while maintaining low adaptation cost, with demonstrated performance advantages over domain-specific models. Specifically, our main contributions are threefold. First, we integrate LoRA with the DINOv3 ViT for seismic denoising and demonstrate that the proposed approach reduces the number of trainable parameters compared with full fine-tuning while achieving performance comparable to or exceeding that of dedicated SFMs. Second, we introduce a TTA mechanism that leverages statistical kurtosis to identify information-rich seismic patches, thereby enabling the model to adapt to previously unseen noise distributions across different field datasets without supervision. Third, we conduct cross-domain benchmarking using field seismic data and DAS-vertical seismic profile (VSP) datasets, which confirms the framework’s strong capability in handling noise variability and domain shifts. Together, these advances establish a scalable and efficient pathway for repurposing pre-trained VFMs in practical geophysical processing. Numerical experiments on both active seismic image data and passive raw DAS-VSP data demonstrate that the proposed framework achieves remarkable trade-off between noise suppression and signal preservation. 

  \section{Problem Formulation and Data Preparation}

  This section presents the formulation of the seismic denoising problem and the preparation of the datasets used in this study. We first define seismic denoising as an inverse problem and describe the learning objective of the proposed framework. We then introduce the datasets constructed for training and evaluation, including both active-source seismic image data and passive, raw DAS-VSP data, followed by the preprocessing procedures applied to ensure consistency across different acquisition conditions. These steps establish the theoretical and experimental foundations for the subsequent development of the proposed framework.
    
  Seismic denoising can be formulated as an inverse problem. Let $X_{obs} \in \mathbb{R}^{H \times W}$ denote the observed (noisy) seismic data, where $H$ and $W$ usually represent the number of time samples (height) and spatial traces (width), respectively. This data is modeled as a superposition of the unknown clean signal $X_{clean}$ and additive noise $\eta$:
  
    \begin{equation}
    X_{obs} = X_{clean} + \eta.\label{eq:noise_model}
    \end{equation}

  Our objective is to learn a mapping function $f_\theta: \mathbb{R}^{H \times W} \to \mathbb{R}^{H \times W}$ parameterized by weights $\theta$, such that the predicted output $\hat{X}_{clean} = f_\theta(X_{obs})$ approximates the ground truth $X_{clean}$ by minimizing a reconstruction loss metric $\mathcal{L}(X_{clean}, \hat{X}_{clean})$. In the context of our proposed framework, the function $f_\theta$ comprises a pre-trained encoder $E_{\phi}$ and a task-specific decoder $D_{\psi}$, parameterized by $\phi$ and $\psi$, respectively, such that the overall parameter set is $\theta = \{\phi, \psi\}$. The primary challenge lies in optimizing these parameters to achieve robust generalization across varying noise distributions without requiring massive labeled datasets for every new survey. Rather than training a network from scratch, as this would necessitate prohibitive amounts of annotated seismic data, we leverage the rich and generalized feature representations already embedded in pre-trained VFMs. Because VFMs possess strong visual priors, adapting them to a new domain is data-efficient. Consequently, to solve this inverse problem and train $\theta$, we require only a small set of paired samples $\{X_{obs}, X_{clean}\}$ (thousands of pairs in our study) to transfer the pre-trained VFMs from natural images to seismic data, which motivates the construction of the datasets described below.

  To facilitate the training and evaluation of the proposed framework, we constructed two distinct datasets representing different acquisition modalities: active seismic image data and passive raw DAS-VSP data.  For the active source data denoising experiments, we utilized the open-source seismic denoising dataset provided by \cite{sheng2025seismic}. This dataset consists of two components designed to test domain generalization. The training subset comprises 2,000 synthetic pairs of clean and noisy seismic sections. The clean data $X_{clean}$ were generated via forward modeling, while the noisy input $X_{obs}$ was created by superimposing random noise with a varying signal-to-noise ratio onto these profiles. The test subset includes 4,000 real-world data pairs extracted from exploration fields. These field samples contain complex coherent noise patterns that differ from the synthetic noise in the training subset, thereby providing a rigorous benchmark for assessing the model's ability to recover $X_{clean}$ under previously unseen noise distributions.

  For the passive seismic denoising experiments, we utilized DAS-VSP data from the Utah FORGE field site \cite[]{lellouch2020comparison}. Data acquisition was conducted in well 78-32 using fiber-optic cables cemented behind the casing. Consequently, the raw observed data $X_{obs}$ are heavily contaminated by optical system noise and coupling noise arising from the cable-borehole interaction. These disturbances appear as coherent zigzag patterns, ringing, and high-amplitude artifacts. To obtain the corresponding ground truth labels $X_{clean}$ for training, we applied the integrated denoising framework described by \cite{dasdenoising2023}, which employs a cascaded filtering strategy designed to suppress various complex noise types and improve the signal-to-noise ratio of DAS data. Following their processing method, we constructed a dataset comprising 194 2D DAS-VSP records, each with original dimensions of $2,000$ time samples $\times$ $960$ channels, encompassing 82 earthquake events and 112 microseismic events.

  To ensure consistent input dimensions for the respective encoders and to mitigate amplitude variations across different acquisition sites, we applied a standardized preprocessing pipeline. First, since the original DAS-VSP records ($2,000 \times 960$) exceed typical network input sizes, we employed a sliding-window method to segment the recordings into smaller patches. Rather than using fixed dimensions, the window sizes and shift steps were dynamically adapted to match the specific pre-training resolution requirements of each utilized VFM. During this segmentation process, minimal overlap was maintained to preserve the continuity of seismic events across patches and to increase the training data volume. Ultimately, this procedure yielded 5,600 training patches derived from the 112 DAS-VSP data pairs containing microseismic events. An independent test set was formed using the remaining 82 DAS-VSP data pairs that capture earthquake events.

  To further address variations in energy levels and noise distributions across datasets, we applied Z-score normalization to all samples. Seismic amplitudes often exhibit large dynamic ranges that can destabilize the training of deep neural networks. Z-score normalization rescales the distribution of seismic amplitudes to have zero mean and unit standard deviation. For a given seismic data patch $X_{patch}$, the normalized sample $X_{norm}$ is calculated as:
  
  \begin{equation}
X_{norm} = \frac{X_{patch} - \mu}{\sigma},\label{eq:zscore}
  \end{equation}

  \noindent where $\mu$ and $\sigma$ represent the mean and standard deviation of the amplitude values within the individual patch $X_{patch}$. This normalization is critical for aligning the noise distributions between the synthetic training data and real-world field data, or cross-site seismic data, thereby facilitating more robust feature extraction during the denoising process. While geometric transformations (e.g., rotation, flipping) are common in computer vision, we avoided additional augmentation for the seismic data. This decision preserves the signal integrity and specific characteristics of seismic wavefronts, preventing the introduction of non-physical artifacts that mislead the network regarding the spatial coherence of geological structures.

  \section{Proposed Parameter-Efficient Framework}
  
  Our proposed framework combines a LoRA-adapted VFM encoder for feature extraction with a lightweight decoder for signal reconstruction. This section first describes the selection of the backbone VFM. It then explains the adaptation strategies, including LoRA-based PEFT and a kurtosis-guided unsupervised TTA approach. Next, the architecture of the asymmetric encoder–decoder network is described. Finally, the training procedure, optimization objectives, and evaluation settings are provided to enable robust and efficient seismic denoising across diverse acquisition conditions.
  
  \subsection{Backbone Vision Foundation Models}
 
  To implement the proposed framework, we first select an appropriate VFM as the backbone encoder for feature extraction. We adopt DINOv3 (Self-Distillation with No Labels v3, follow-up of DINOv2 \cite[]{oquab2023dinov2})  as the backbone encoder due to its strong global-local feature extraction capabilities and its ability to learn robust semantic features such as object boundaries and textures rather than memorizing pixel-level values. To validate this choice, we benchmark DINOv3 against two representative encoder architectures: SFM, a domain-specific Transformer pre-trained on millions of seismic patches that offers domain-specific priors, and SwinV2, a hierarchical ViT pre-trained on ImageNet-22k dataset \cite[]{ridnik2021imagenet}.

  The DINOv3 architecture follows the ViT design, which divides input images into a sequence of non-overlapping patches. In our study, the input seismic data $X_{obs}$ is first divided into a sequence of flattened patches and projected into a latent embedding sequence $Z \in \mathbb{R}^{N \times d}$, where $N$ denotes the total number of patches and $d$ represents the embedding dimension. These embeddings are processed by a stack of multi-head self-attention layers, which allows the model to capture long-range dependencies through a global receptive field. The core attention operation \cite[]{vaswani2017attention} calculates the relevance of each patch with respect to all other patches via the following equation:
  
    \begin{equation}
      \text{Attention}(Z) = \text{softmax}\left(\frac{(ZW^Q)(ZW^K)^T}{\sqrt{d_k}}\right)(ZW^V), 
      \label{eq:attention}
    \end{equation}
  
  \noindent where $W^Q, W^K, W^V \in \mathbb{R}^{d \times d_k}$ represent the dense projection weight matrices for the Query (Q), Key (K), and Value (V), respectively, and $d_k$ is the hidden dimension of each attention head. These matrices transform the seismic embeddings $Z$ into a feature space where signal coherence can be effectively distinguished from noise.

  \subsection{Adaptation Strategies}
  
  Adapting these high-dimensional projection matrices ($W^Q, W^K, W^V$) to data-set-specific textures typically requires updating the entire encoder parameter set $\phi$. This full fine-tuning process is computationally expensive and prone to overfitting, particularly given the limited scale of labeled seismic datasets. To address this challenge efficiently, we implement PEFT via LoRA, targeting these attention projection weights. LoRA freezes the pre-trained weights $W_0$ (e.g., the original $W^Q$) and injects trainable low-rank matrices $A \in \mathbb{R}^{d \times r}$ and $B \in \mathbb{R}^{r \times d}$ into the projection path, where $d$ represents the original embedding dimension and $r$ is the low-rank bottleneck dimension. Under the assumption that the adaptation to seismic noise patterns lies in a low intrinsic-rank subspace $r \ll d$, the forward pass for a specific projection layer acting on the seismic input features $x \in Z$ is expressed as:
  
  \begin{equation}
      h = W_0 x + \Delta W x = W_0 x + BAx. \label{eq:lora}
  \end{equation}

  We initialize $A$ with random Gaussian noise and $B$ with zeros to ensure $\Delta W = 0$ at the start of training, thereby preserving the original feature extraction capability of DINOv3. Specifically, we apply LoRA to the Query, Key, Value, and Output projection matrices with a designated rank $r$ and a scaling factor $\alpha$. The rank $r$ defines the inner dimensionality of the low-rank matrices, serving as an information bottleneck, restricting the adaptation space to mitigate overfitting on the downstream task. Concurrently, the scaling factor $\alpha$ dictates the magnitude of the impact that the low-rank updates $\Delta W$ have on the original frozen weights, as the learned activations are scaled by $\alpha/r$. This configuration results in a substantial reduction in trainable parameters compared to full fine-tuning, thereby preventing catastrophic forgetting of the general vision features learned during pre-training.

  Finally, to address domain shifts and enhance denoising performance on unseen real-world data, we incorporate a kurtosis-guided unsupervised TTA phase. This approach is motivated by the observation that useful seismic signals exhibit sparsity and non-Gaussianity in the time-space domain, resulting in higher kurtosis values ($\kappa$) than those associated with the Gaussian distributions typically observed in background noise. This statistical property forms the core prior of the kurtosis-based sample selection strategy proposed for unsupervised DAS-VSP denoising \cite[]{cui2025unsupervised}. For a given input seismic patch $x_{patch} \subset X_{obs}$, the kurtosis $\kappa$, which measures the peakedness of the amplitude distribution, is defined as:
  
  \begin{equation}
      \kappa(x_{patch}) = \frac{\mathbb{E}[(x_{patch} - \mu)^4]}{\sigma^4},\label{eq:kurtosis}
  \end{equation}

  \noindent where $\mu$ and $\sigma$ are the mean and standard deviation of the patch amplitudes, respectively. During the TTA phase, we normalize the raw data $X_{obs}$ using Z-scores (cf. Eq. \ref{eq:zscore}) and segment it into $224 \times 224$ patches. Subsequently, the algorithm calculates $\kappa$ for every patch to differentiate information-rich regions from noise-dominated areas in $X_{obs}$.

  \begin{figure}[htbp]
    \centering
    \includegraphics[width=1.0\textwidth]{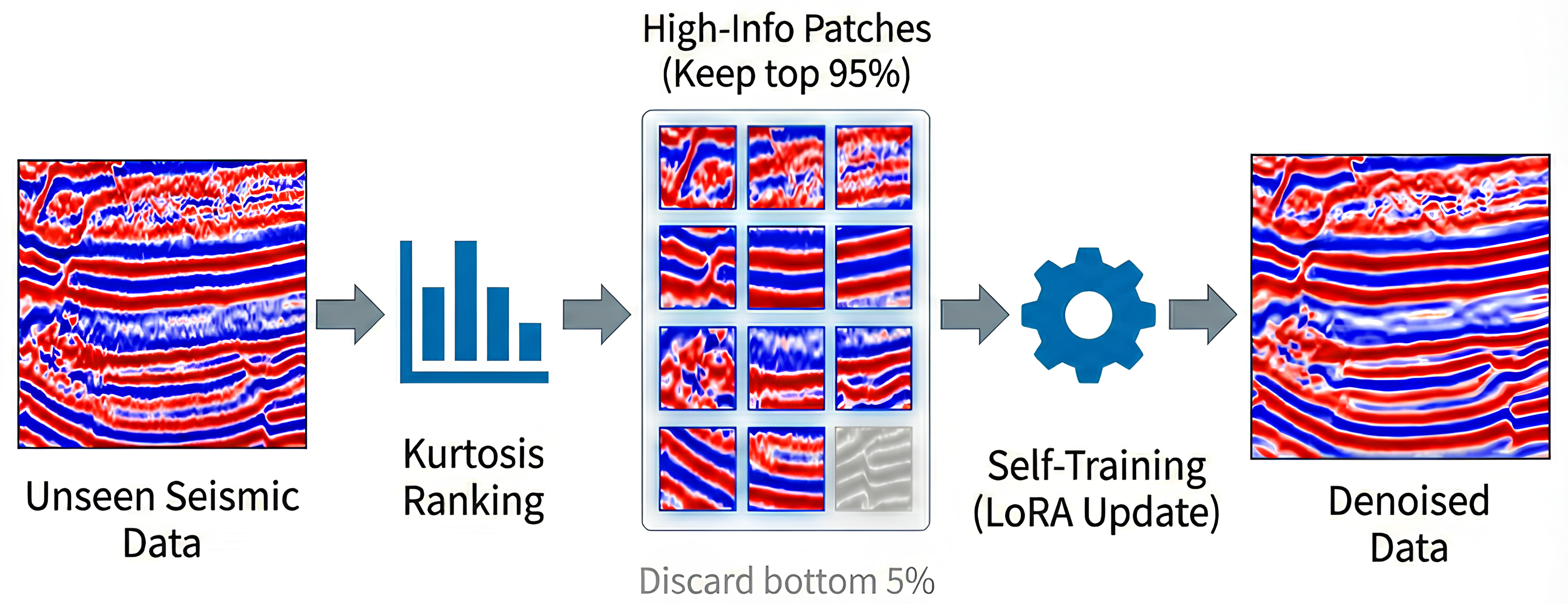}
    \caption{The proposed kurtosis-guided TTA workflow. The raw unseen seismic data are first segmented into patches. These patches are ranked by their kurtosis values; the top 95\% of high-information patches are selected to self-train the LoRA parameters, while the bottom 5\% of noise-dominated patches are discarded to prevent overfitting to background noise.}
    \label{fig:kurtosis_workflow}
  \end{figure}

  As illustrated in Figure \ref{fig:kurtosis_workflow}, we rank the patches based on their kurtosis values. A threshold is applied to retain the top $K\%$ (e.g., 95\%) of patches with the highest kurtosis, as these regions contain more significant waveform information that is essential for effective feature adaptation \cite[]{cui2025unsupervised}. These selected patches are then used to self-calibrate the LoRA parameters $A$ and $B$ via an unsupervised reconstruction loss, allowing the model to capture local signal textures specific to the new domain. Conversely, the bottom 5\%, which correspond to patches dominated by random background noise, are discarded to prevent the model from learning irrelevant or misleading patterns. Once optimized, the adapted model processes the full input to generate denoised data with preserved structural integrity.

  \subsection{Framework Architecture}

\begin{figure}
    \centering
    \includegraphics[width=1.0\textwidth]{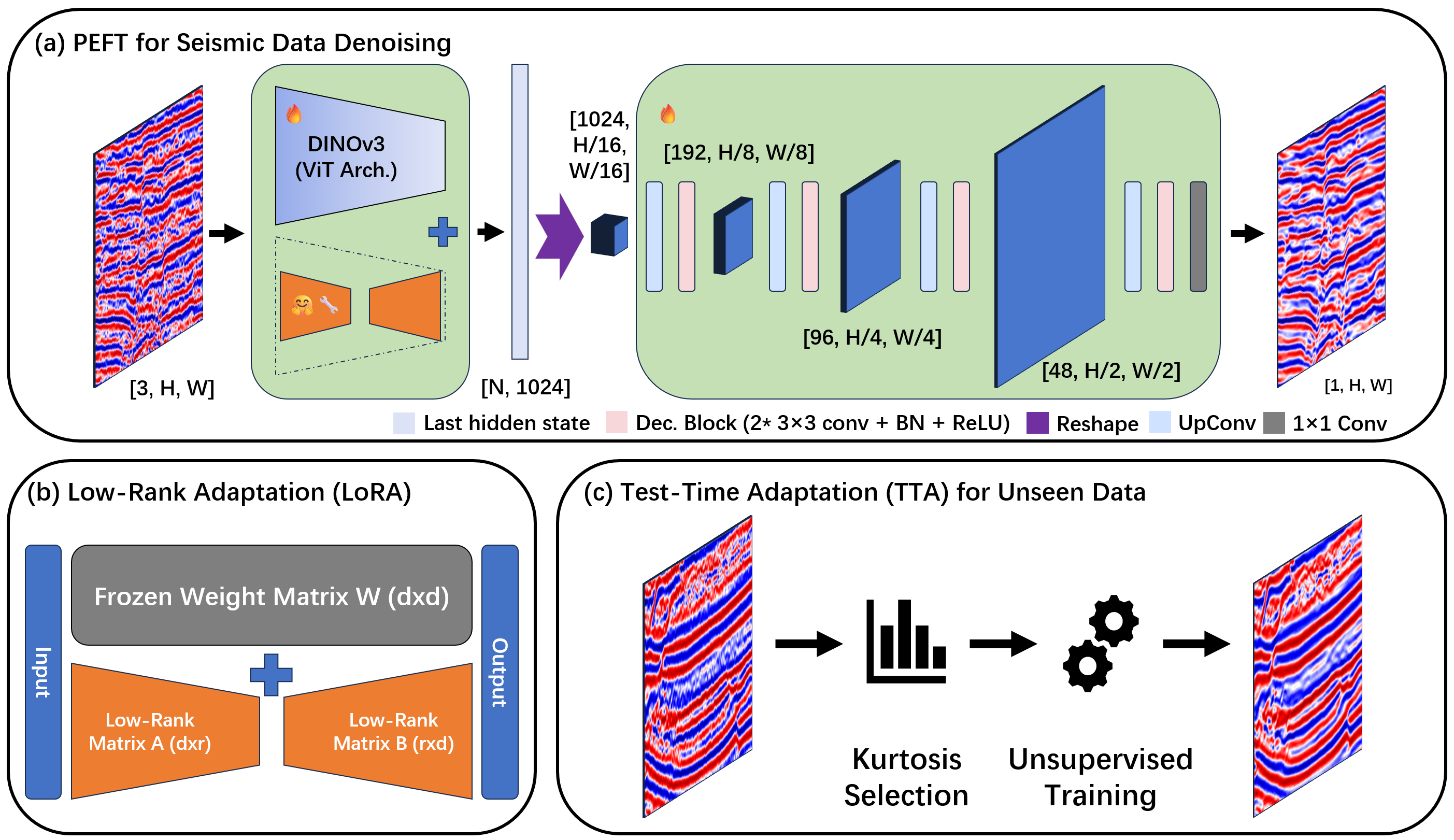}
    \caption{Overview of the proposed framework and its key components. (a) The encoder-decoder module, where seismic data is fed into a encoder (e.g. DINOv3) adapted via LoRA. The extracted latent features are reshaped and processed by a lightweight decoder to produce the denoised output. (b) The LoRA module, which injects trainable low-rank matrices ($A$ and $B$) parallel to the frozen pre-trained weight matrix $W$ for PEFT. (c) The kurtosis-guided TTA workflow: for unseen seismic data, real seismic data is first filtered via kurtosis selection to retain high-information patches, then refined through unsupervised learning, and finally reconstructed into denoised data. }
    \label{framework}
  \end{figure}

  This subsection describes the overall architecture of the proposed framework, which employs an asymmetric encoder-decoder design. As illustrated in Figure \ref{framework}a, it integrates a LoRA-adapted VFM encoder for feature extraction with a streamlined decoder for signal reconstruction. 

  Specifically, the left section of Figure \ref{framework}a depicts the encoder module, which utilizes a LoRA-adapted VFM backbone (in this case, DINOv3) to process input seismic patches and generate a hierarchy of latent feature maps. Because pre-trained VFMs like DINOv3 expect three-channel RGB inputs, the raw single-channel seismic patches ($1 \times 224 \times 224$) are first expanded and duplicated across the channel dimension to form $3 \times 224 \times 224$ inputs. To bridge the ViT's 1-D sequence output with the 3D spatial requirements of the decoder, we introduce a tailored token extraction mechanism. Modern VFMs prepend auxiliary tokens to their sequences; thus, the raw output comprises a class token, multiple register tokens which are designed to suppress artifacts in attention maps, and the actual spatial patch tokens. Taking DINOv3 as an example, given a patch size of 16, a $3 \times 224 \times 224$ input patch yields a sequence of 201 tokens, including 1 class token, 4 register tokens and 196 patch tokens. 
  
  To accurately reconstruct the spatial hierarchy, our module bypasses the initial $1 + 4$ auxiliary tokens and retains only the relevant patch tokens. These extracted tokens are subsequently permuted and reshaped back into a 3D spatial feature map (e.g., $1024 \times \frac{H}{16} \times \frac{W}{16}$), which can be directly processed by the decoder. To efficiently adapt the pre-trained VFMs to the seismic domain without incurring prohibitive computational costs, the encoder incorporates LoRA, as detailed in Figure \ref{framework}b. During the forward pass, the input features are processed simultaneously by both the frozen weights and the low-rank branch, with their outputs subsequently summed (as described in Eq. \ref{eq:lora}). This strategy allows the network to learn domain-specific seismic representations with a minimal number of trainable parameters, reducing memory overhead while preserving the generalized feature extraction capability learned during pre-training.

  Subsequently, these latent representations are processed by a lightweight decoder designed to reconstruct high-resolution seismic images. In structure, this decoder comprises a series of upsampling blocks, where each block utilizes a $2 \times 2$ Transposed Convolution with a stride of 2 to double the spatial resolution, followed by Batch Normalization and a ReLU activation function. At the end of the decoding process, the resulting multi-channel output is averaged along the channel dimension to collapse the data back into a single-channel ($1 \times 224 \times 224$) seismic patch, matching the dimension of the original seismic patch.
  
  A characteristic of this architecture, diverging from standard U-Net designs, is the omission of long skip connections between the encoder and decoder. Although skip connections typically assist in preserving high-frequency details by linking inputs directly to outputs, they are counterproductive in this specific denoising context as they permit noise to bypass the filtering bottleneck. Therefore, by compelling all information to traverse the bottleneck, the proposed framework ensures that the final reconstruction is derived from the clean semantic features identified by the encoder. 
  
  Finally, as shown in Figure \ref{framework}c, this architectural backbone is coupled with a kurtosis-guided TTA workflow, which filters raw data to retain high-information patches for unsupervised refinement, thereby ensuring robustness on unseen seismic surveys.

  \subsection{Training and Evaluation Setup}
  
  All experiments in this study were conducted on a single NVIDIA Tesla V100 (32 GB) GPU utilizing the PyTorch framework. The training workflow comprises two distinct phases: a supervised PEFT stage and an unsupervised TTA stage during inference. To accommodate the architectural differences among the utilized VFMs, we adapted the input resolutions and sliding window parameters accordingly, as summarized in Table~\ref{tab:parameters}. 

\begin{table}[htbp]
  \centering
  \caption{Part of hyperparameter configurations for the evaluated VFMs during the PEFT and TTA stages. For the shift step, values appended with (known) and (unseen) denote the settings applied when inferring from data with known or similar features versus unknown data, respectively.}
  \label{tab:parameters}\renewcommand*{\arraystretch}{1.1}
  \begin{tabular}{|l|c|c|}
    \hline
    \textbf{Hyperparameter} & \textbf{SFM \& DINOv3} & \textbf{SwinV2} \\
    \hline
    Input Size & $224 \times 224$ & $256 \times 256$ \\
    \hline
    \multirow{2}{*}{Shift Step} & $221 \times 221$ (known) & $253 \times 253$ (known) \\
     & $28 \times 28$ (unseen) & $32 \times 32$ (unseen) \\
     \hline
    \multirow{2}{*}{LoRA Target Modules} & \texttt{q\_proj}, \texttt{k\_proj},  & \texttt{q\_proj}, \texttt{k\_proj},  \\
     & \texttt{v\_proj}, \texttt{o\_proj} & \texttt{v\_proj}, \texttt{o\_proj} \\
    \hline
    LoRA Rank ($r$) & \multicolumn{2}{c|}{16} \\
    \hline
    LoRA Scaling Factor ($\alpha$) & \multicolumn{2}{c|}{64} \\
    \hline
    LoRA Dropout Rate & \multicolumn{2}{c|}{0.1} \\
    \hline
  \end{tabular}
\end{table}
  
  \noindent All models maintain the same settings as in Table~\ref{tab:parameters} during both the training and inference phases. During the supervised phase, we optimize the model parameters using the AdamW optimizer \cite[]{loshchilov2017decoupled}. To stabilize the early stages of training and prevent premature convergence, we incorporate a learning rate warmup phase \cite[]{goyal2017accurate}, followed by a cosine annealing learning rate schedule \cite[]{loshchilov2016sgdr} to smoothly decay the learning rate as training progresses. Regarding the PEFT configuration, we utilize the Hugging Face PEFT library \cite[]{peft} to apply LoRA. As detailed in Table~\ref{tab:parameters}, this adaptation targets the specific projection modules of the Transformer blocks across all backbones. After the fine-tuning stage, the TTA involves unsupervised fine-tuning for additional epochs during inference. For this adaptation, we reduce the learning rate to ensure stable adaptation to unseen noise distributions without erasing the learned semantic priors. More detailed hyperparameter settings and training schedules are provided in the following experimental section.

  To train the model effectively, we minimize a composite loss function $\mathcal{L}$ designed to balance pixel-wise fidelity with structural preservation. This objective function is formulated as:
  
  \begin{equation}
      \mathcal{L} = (1 - \lambda) \cdot \mathcal{L}_{\text{MSE}} + \lambda \cdot \mathcal{L}_{\text{MS-SSIM}}.
      \label{eq:loss}
  \end{equation}
 
  In this equation, the first component is the Mean Squared Error (MSE), calculated as: 

  \begin{equation}
      \mathcal{L}_{\text{MSE}} = \frac{1}{N} \sum_{i=1}^N (X_{clean, i} - \hat{X}_{clean, i})^2,\label{eq:mse}
  \end{equation}

  \noindent where $X_{clean}$ represents the clean ground truth and $\hat{X}_{clean}$ denotes the denoised output. This term ensures that the amplitudes of the denoised data match those of the ground truth. Simultaneously, to address structural fidelity, the loss incorporates the Multi-Scale Structural Similarity Index (MS-SSIM) \cite[]{wang2003multiscale}. The MS-SSIM evaluates image quality progressively across $M$ scales, capturing hierarchical structural details. It is formulated as:
  
  \begin{align}
    \text{MS-SSIM}(X_{\text{clean}}, \hat{X}_{\text{clean}}) &= [l_M(X_{\text{clean}}, \hat{X}_{\text{clean}})]^{\alpha_M} \nonumber\\
    &\times \prod_{j=1}^{M} [c_j(X_{\text{clean}}, \hat{X}_{\text{clean}})]^{\beta_j} [s_j(X_{\text{clean}}, \hat{X}_{\text{clean}})]^{\gamma_j},\label{eq:msssim}
  \end{align}
 
 \noindent where $l_M$, $c_j$, and $s_j$ denote the luminance, contrast, and structure comparison measures at scale $j$, respectively. Because the MS-SSIM is a similarity metric bounded between 0 and 1 (where a value of 1 indicates perfect structural fidelity), it represents a maximization objective. However, gradient-based neural network optimizers are designed to minimize a cost function. To integrate this metric into our proposed framework, we reformulate it into a structural dissimilarity loss: 
  
  \begin{equation}
    \mathcal{L}_{\text{MS-SSIM}} = 1 - \text{MS-SSIM}(X_{\text{clean}}, \hat{X}_{\text{clean}}),
  \end{equation}
  
  \noindent which converts the metric into a penalty for structural distortions, forcing the loss to converge towards zero as the prediction approaches the ground truth. We set the balancing coefficient $\lambda$ to 0.5, a configuration that assigns equal importance to the competing objectives of amplitude restoration and structural preservation for seismic data.

  To validate the efficacy of our proposed framework, we benchmark it against two other representative VFMs with different numbers of model parameters. We evaluate SFM-Large (hereafter referred to as SFM) which is developed based on ViT-Large, with approximately 300 million model parameters, and SwinV2-B (hereafter referred to as SwinV2) with approximately 88 million  model parameters, testing them under both full fine-tuning and LoRA-adapted settings. To isolate the benefits of the proposed efficiency strategy, we compare our proposed framework against a full fine-tuning DINOv3 (DINOv3 ViT-S/16 with approximately 21.6 million model parameters) where all parameters are updated without low-rank constraints. 
  
  The quantitative assessment of denoising performance employs two distinct metrics tailored to the specific characteristics of the datasets. For conventional active seismic image data, we utilize the MS-SSIM index to evaluate structural consistency across varying resolutions. In this context, higher MS-SSIM values indicate superior preservation of structural integrity and texture. Conversely, for the DAS-VSP data, we employ the local similarity (LS) attribute \cite[]{fomel2007local, chen2015random}. This metric measures the orthogonality between the denoised signal section and the removed noise section within a local neighborhood. Consequently, lower LS values signify minimal signal leakage into the noise residual, reflecting a superior separation of the microseismic events from the background noise.

  \section{Experiments and Results}

  In this section, we apply the proposed framework to both active seismic image data and passive raw DAS-VSP data to evaluate its denoising performance across diverse dataset types. To further assess its generalization capability, we conduct additional tests on two unseen real-world field datasets.

  To validate the effectiveness of the proposed framework, we benchmark different backbone VFMs used as encoders and compare our approach against several state-of-the-art methods. For active seismic denoising, we employ the pre-trained SFM fine-tuned on the same dataset as a strong domain-specific baseline. Meanwhile, for the generalization experiments in active seismic image data denoising, we also include an advanced self-supervised method, Noise2Void (N2V) \cite[]{birnie2021potential}, which utilizes a blind-spot network to predict the noise-free value of a central sample from its surroundings, based on the assumption that random noise is statistically independent across samples. 
  
  For passive DAS-VSP denoising, we compare against two advanced baselines: DAS-N2N \cite[]{lapins2024n2n}, a weakly supervised model that leverages the pairwise properties of DAS data as a supervision signal; and a lightweight, unsupervised method based on an Attention-Based Deep Image Prior (ABDIP) \cite[]{cui2025unsupervised}.

  \subsection{Active Seismic Denoising Results}

  We evaluate five configurations on the pre-processed 2D active seismic image dataset: full fine-tuning SFM (SFM Fine-tune), full fine-tuning (DINOv3 Fine-tune), and our proposed framework (which integrates LoRA-based PEFT with kurtosis-guided unsupervised TTA, denoted as PEFT+TTA) applied to SFM, SwinV2, and DINOv3. This comparison enables a systematic assessment of denoising performance across different backbone encoders and adaptation strategies on real-world data. The training hyperparameters for all configurations are detailed in Tables~\ref{tab:parameters} and \ref{tab:hyperparameters}, and the corresponding evaluation results on the validation dataset are presented in Table~\ref{tab:results}.

  \begin{table}[htbp]
    \centering
    \caption{Main training hyperparameters and configuration settings for the evaluated models (SFM, SwinV2, DINOv3) with different trainable parameters (T.Params) under different adaptation strategies (full Fine-tune, and PEFT). For the PEFT+TTA configurations, the values appended with (TTA) denote the specific settings applied during the Test-Time Adaptation phase.}
    \label{tab:hyperparameters}\renewcommand*{\arraystretch}{1.2}
    \begin{tabular}{|l|c|c|c|c|c|}
      \hline
      \multirow{2}{*}{\textbf{Settings}} & \multicolumn{2}{c|}{\textbf{Fine-tune}} & \multicolumn{3}{c|}{\textbf{PEFT+TTA}} \\
      \cline{2-6}
       & \textbf{SFM} & \textbf{DINOv3} & \textbf{SFM} & \textbf{SwinV2} & \textbf{DINOv3} \\
      \hline
      T.Params (M) & 306.38 & 22.87 & 5.96 & 6.66 & 1.87  \\
      \hline
      Batch Size & \multicolumn{5}{c|}{24} \\
      \hline
      Base LR & \multicolumn{2}{c|}{$6.4 \times 10^{-4}$} & \multicolumn{3}{c|}{$6.0 \times 10^{-4}$; $1.0 \times 10^{-4}$ (TTA)}  \\
      \hline
      LR Schedule & \multicolumn{5}{c|}{cosine}  \\
      \hline
      Training Epochs & \multicolumn{2}{c|}{100} & \multicolumn{3}{c|}{50 + 50 (TTA)}  \\
      \hline
    \end{tabular}
\end{table}

  \begin{table}[htbp]
    \centering
    \caption{Comparison of denoising performance on the active seismic field validation dataset. The evaluation metrics include MS-SSIM (denoised vs. original) and MS-SSIM-R (noise vs. original). Note that $\uparrow$ ($\downarrow$) indicates that a larger (smaller) value represents better performance, with the best results shown in bold.}
    \label{tab:results}\renewcommand*{\arraystretch}{1.2}
    \begin{tabular}{|lcc|} 
      \hline
      \textbf{Method} & \textbf{Avg. MS-SSIM $\uparrow$} & \textbf{Avg. MS-SSIM-R $\downarrow$} \\
      \hline
      SFM Fine-tune & 0.9169 & 0.5206 \\
      \hline
      DINOv3 Fine-tune & 0.8962 & 0.4905 \\
      \hline
      SFM (PEFT+TTA) & \textbf{0.9885} & \textbf{0.4146} \\
      SwinV2 (PEFT+TTA) & 0.8150 & 0.4482 \\
      DINOv3 (PEFT+TTA) & 0.9556 & 0.4730 \\
      \hline
    \end{tabular}
  \end{table}

    Since noise-free ground truth labels are unavailable for field seismic data, we adopt the evaluation metrics established by \cite{sheng2025seismic}. Specifically, we calculate the MS-SSIM between the denoised result and the original seismic data, together with the MS-SSIM between the noise and the original seismic data, which we denote as MS-SSIM-R. Effective denoising is characterized by high structural similarity between the denoised result and the original data (i.e., high MS-SSIM), along with low structural similarity between the removed noise and the original seismic data (i.e., low MS-SSIM-R). 
    
    The results in Table \ref{tab:results} demonstrate clear advantages of the proposed framework which employs LoRa-based PEFT and kurtosis-guided unsupervised TTA. For domain-specific models, applying our proposed framework to SFM achieves the absolute best overall performance, reaching an average MS-SSIM of 0.9885 and an MS-SSIM-R of 0.4146. It outperforms the SFM Fine-tune baseline while updating only a fraction of the trainable parameters. 
    
    For general VFMs that lack seismic priors, the DINOv3 Fine-tune establishes a strong baseline. Notably, DINOv3 adapted via our proposed framework achieves outstanding performance (MS-SSIM: 0.9556, MS-SSIM-R: 0.4730). It not only surpasses the DINOv3 Fine-tune but also outperforms the full fine-tuning domain-specific model (SFM Fine-tune). This demonstrates our proposed framework's exceptional efficiency, indicating that our adaptation strategy bridges the domain gap and enables cross-domain VFMs to rival or even exceed full fine-tuning specialized models. 
    
    The SwinV2 model adapted through our proposed framework exhibits lower signal preservation, with an MS-SSIM of 0.8150, while still maintaining strong noise separation capabilities, achieving an MS-SSIM-R of 0.4482 that outperforms both full fine-tuning baselines. This performance pattern suggests that the domain shift between general visual representation learning and specialized seismic denoising remains significant. Architectural characteristics, such as SwinV2's window-based local attention versus DINOv3's global representation, may dictate how effectively a VFM preserves the continuity of complex seismic wavefields.

    \begin{figure}[htbp]
      \centering
      \includegraphics[width=1.0\textwidth]{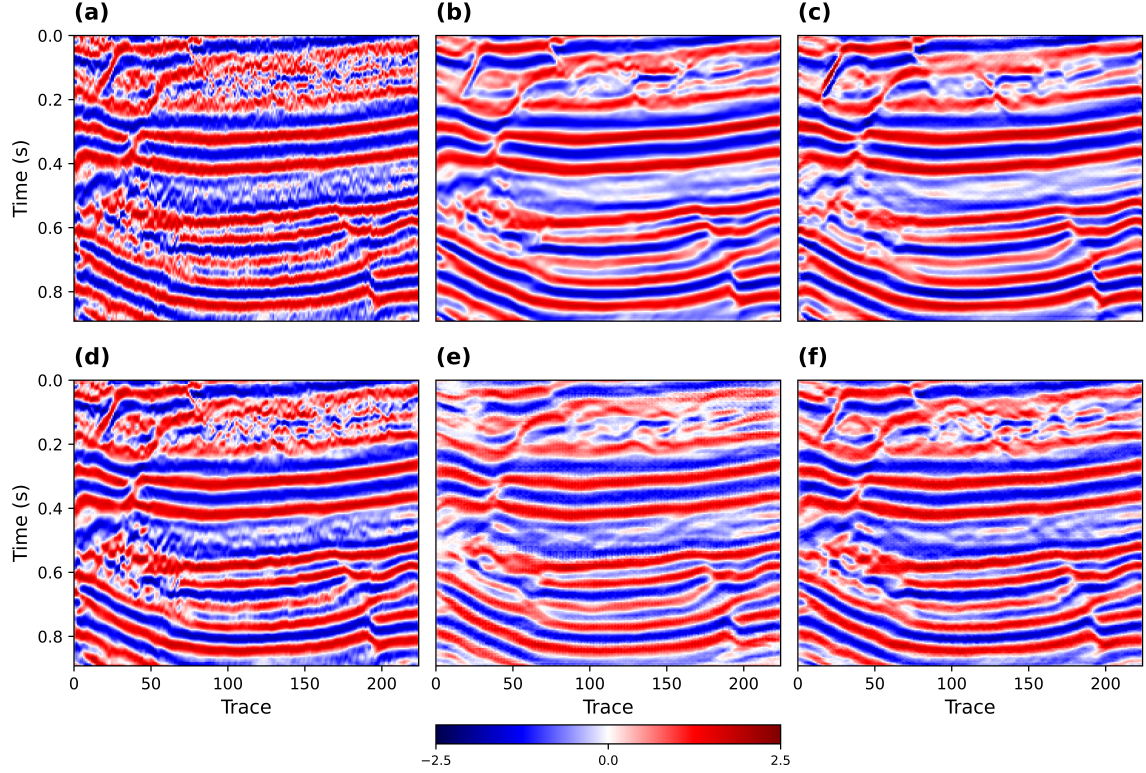}
      \caption{Visual comparison of denoised results across the five configurations on a challenging validation sample. (a) Raw data. Denoised results via: (b) SFM Fine-tune (MS-SSIM = 0.5843), (c) DINOv3 Fine-tune (MS-SSIM = 0.5303), (d) SFM with PEFT+TTA (MS-SSIM = 0.9931), (e) SwinV2 with PEFT+TTA (MS-SSIM = 0.9595), and (f) DINOv3 with PEFT+TTA (MS-SSIM = 0.9799). Our proposed PEFT+TTA frameworks (d and f) demonstrate superior structural recovery and continuity, especially d.}
      \label{fig:visual_comparison}
    \end{figure}

    \begin{figure}[htbp]
      \centering
      \includegraphics[width=1.0\textwidth]{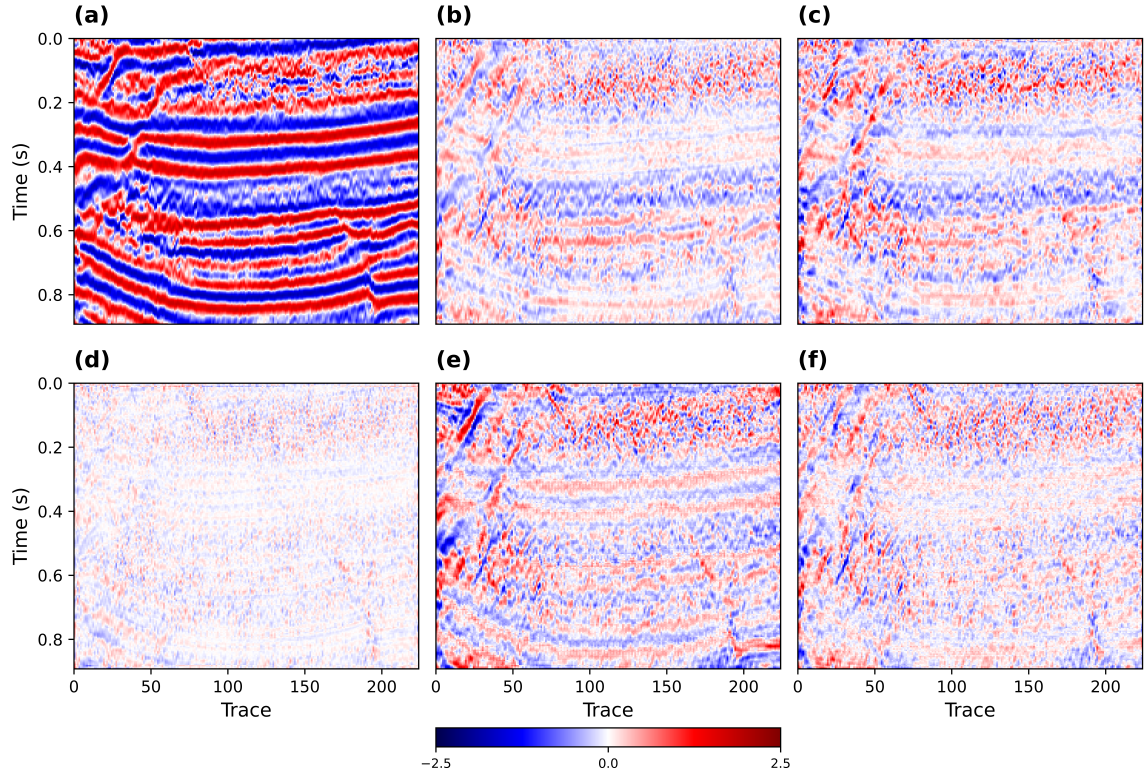}
      \caption{Visual comparison of the removed noise profiles corresponding to Figure \ref{fig:visual_comparison}. (a) Raw data. Extracted noise via: (b) SFM Fine-tune (MS-SSIM-R = 0.4993), (c) DINOv3 Fine-tune (MS-SSIM-R = 0.4931), (d) SFM with PEFT+TTA (MS-SSIM-R = 0.4003), (e) SwinV2 with PEFT+TTA (MS-SSIM-R = 0.5183), and (f) DINOv3 with PEFT+TTA (MS-SSIM-R = 0.4423). Lower MS-SSIM-R indicates less signal leakage. }
      \label{fig:visual_comparison_noise}
    \end{figure}

    Visual comparisons further corroborate these quantitative findings. We selected a challenging sample from the validation set to evaluate the denoised data (Figure~\ref{fig:visual_comparison}) and the corresponding removed noise (Figure~\ref{fig:visual_comparison_noise}) across all five configurations. Consistently with the average numerical metrics, SFM and DINOv3 adapted via our proposed framework effectively remove random noise while maximally preserving the underlying seismic structural information, achieving remarkable MS-SSIM scores of 0.9931 and 0.9799 on this specific sample, respectively. The full fine-tuning models (Figures~\ref{fig:visual_comparison}b and \ref{fig:visual_comparison}c) perform inadequately on this complex sample, showing visible limitations in recovering the finest continuous features and yielding much lower MS-SSIM scores (0.5843 and 0.5303). 
    
    In contrast, while the adapted SwinV2 reconstructs a relatively clean image (Figure~\ref{fig:visual_comparison}e), its corresponding noise profile (Figure~\ref{fig:visual_comparison_noise}e) shows that a noticeable amount of coherent seismic events has leaked into the removed noise section, resulting in the highest MS-SSIM-R of 0.5183 among the PEFT methods, which aligns with the aforementioned architectural limitations of window-based local attention. Ultimately, the separated noise plots of our adapted DINOv3 and SFM models (Figure~\ref{fig:visual_comparison_noise}d and \ref{fig:visual_comparison_noise}f) exhibit unstructured, random characteristics with the lowest MS-SSIM-R values (0.4003 and 0.4423). They transfer learned representations to unseen real-world field data, striking an optimal balance between noise reduction and high-fidelity signal preservation.

  \subsection{Passive DAS-VSP Denoising Results}

  For the DAS-VSP denoising task, we evaluate five configurations on the pre-processed Utah FORGE dataset: DAS-N2N, ABDIP, and our proposed framework utilizing PEFT with SFM, SwinV2, and DINOv3 encoders. Given the lack of noise-free ground truth labels for field DAS-VSP data, we employ the LS metric to quantify signal leakage across the different methods, where a lower value indicates better signal preservation. Except for DAS-N2N and ABDIP, which retain the hyperparameter settings from their original work, the settings for our proposed framework are detailed in Tables~\ref{tab:parameters} and \ref{tab:hyperparameters}. The evaluation results are presented in Table \ref{tab:das_results}.

  \begin{table}
    \centering
    \caption{Comparison of denoising performance on the Utah FORGE DAS-VSP dataset. The table presents the average LS values for DAS-N2N, ABDIP, and our proposed framework with SFM, SwinV2, and DINOv3 encoders, where lower LS values indicate less signal leakage. Moreover, the authors do not normalize the input data in the ABDIP work, and we also retain this option (w/o Norm.) in our experiment. Note that $\downarrow$ indicates that a smaller value represents better performance, with the best results shown in bold.}
    \label{tab:das_results}\renewcommand*{\arraystretch}{1.2}
    \begin{tabular}{|lc|} 
      \hline
      \textbf{Method} & \textbf{Avg. Local Similarity (LS) $\downarrow$} \\
      \hline
      DAS-N2N & \textbf{0.086} \\
      \hline
      ABDIP (w/o Norm.) & 0.127\\
      \hline
      SFM (PEFT+TTA) & 0.167 \\
      SwinV2 (PEFT+TTA) & 0.117 \\
      DINOv3 (PEFT+TTA) & 0.109 \\
      \hline
    \end{tabular}
  \end{table}

    As shown in Table \ref{tab:das_results}, the dedicated DAS-N2N method achieves the best overall performance across the dataset with the lowest average LS of 0.086. However, our proposed framework adapted with the DINOv3 encoder also demonstrates competitive capabilities, securing the second-best average LS of 0.109. This indicates that DINOv3's dense, global feature extraction is effective at capturing and separating the unique noise patterns in DAS-VSP data, even surpassing the average performance of the specialized ABDIP framework (0.127). Interestingly, the domain-specific SFM yields the highest signal leakage with an average LS of 0.167, which stems from a domain mismatch. SFM is primarily pre-trained on active seismic recordings, causing its learned domain priors to struggle against the highly specific optical and coupling noise characteristics of DAS-VSP data. In this scenario, the universal visual textures captured by general VFMs (like DINOv3, and SwinV2 at 0.117) prove to be more robust and adaptable than mismatched domain-specific priors.

  \begin{figure}
    \centering
    \includegraphics[width=0.85\textwidth]{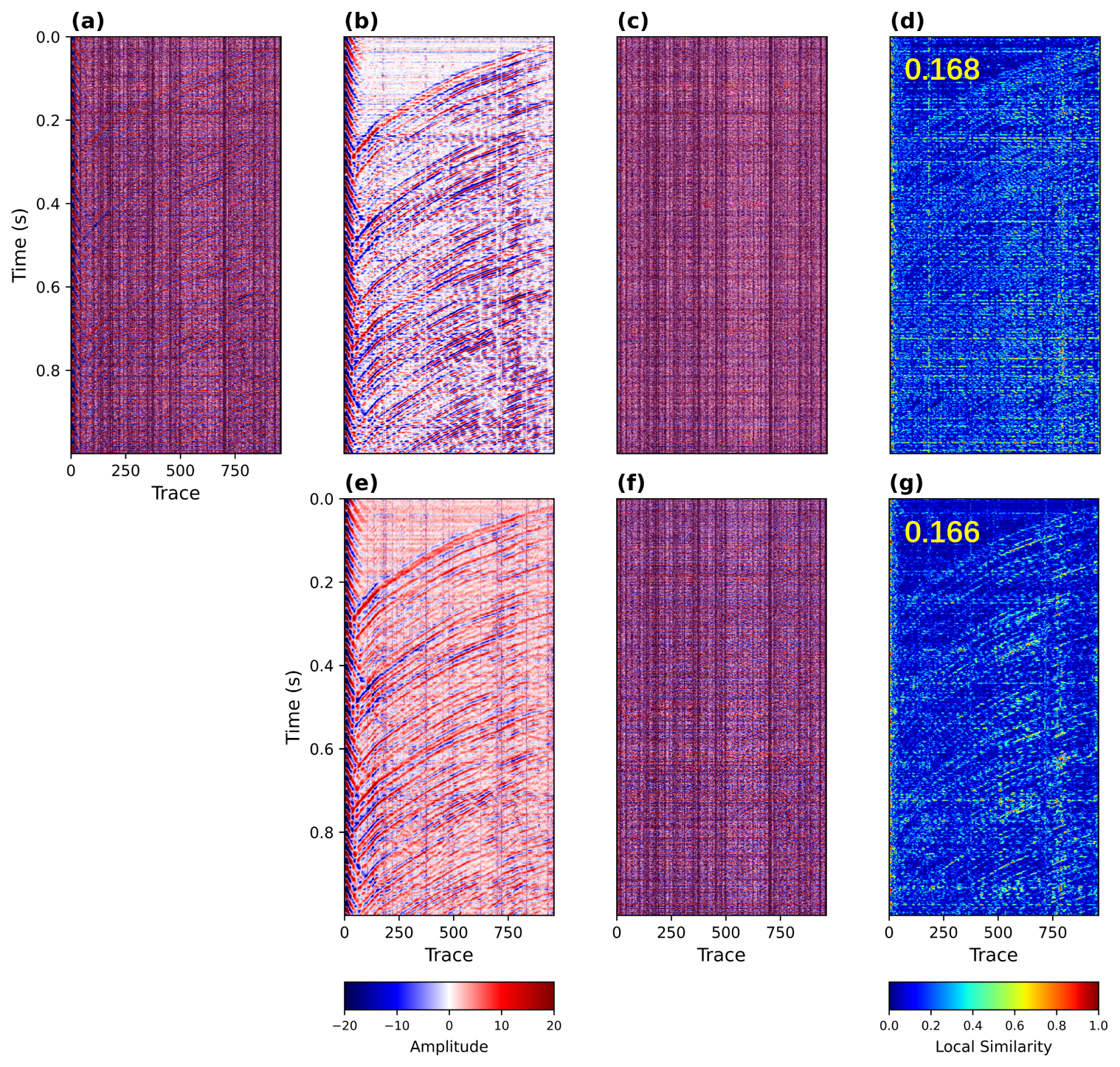}
    \caption{Visual comparison of the denoising results achieved by two state-of-the-art baselines (DAS-N2N and ABDIP) on a representative DAS-VSP sample. (a) Raw data. (b-d) Denoised result, removed noise, and LS map for DAS-N2N (LS = 0.168). (e-g) Results for ABDIP (LS = 0.166). Warmer colors in the LS map indicate higher signal leakage.}
    \label{fig:das_visual1}
  \end{figure}

    \begin{figure}
    \centering
    \includegraphics[width=0.85\textwidth]{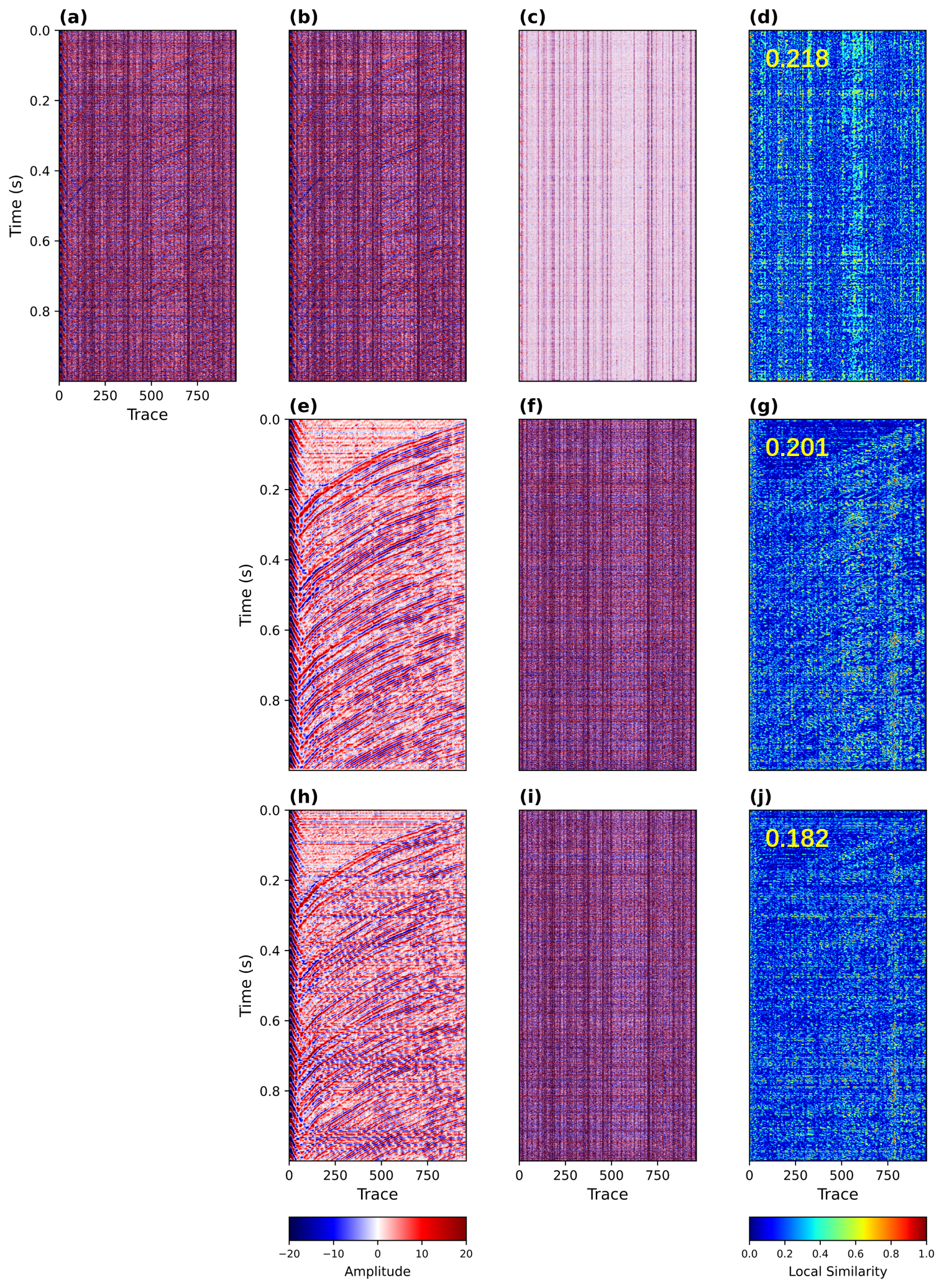}
    \caption{Denoising results of our three adapted models evaluated on the representative DAS-VSP sample. (a) Raw data. (b-d) Results for our SFM-adapted model (LS = 0.218). (e-g) Results for our SwinV2-adapted model (LS = 0.201). (h-j) Results for our DINOv3-adapted model (LS = 0.182).}
    \label{fig:das_visual2}
  \end{figure}

  Figures \ref{fig:das_visual1} and \ref{fig:das_visual2} display the denoising results for a single representative  sample selected from the validation set. The sample is characterized by strong noise, which will lead to incorrect first-arrival picking during subsequent processing. On this specific sample, the visual and quantitative trends slightly deviate from the dataset averages. For this individual profile, the unsupervised ABDIP method achieves the best visual separation with the lowest LS of 0.166, cleanly suppressing noise while preserving continuous wavefields. The dedicated weakly-supervised DAS-N2N method and our proposed framework (with DINOv3) perform comparably well, yielding LS scores of 0.168 (Figure \ref{fig:das_visual1}d) and 0.182 (Figure \ref{fig:das_visual2}j), respectively. Both methods effectively suppress the complex coupling noise, though slight signal leakage can be observed in their LS maps. Conversely, the SFM-adapted model (LS = 0.218) struggles to cleanly separate the zigzag optical noise from signals, validating its higher LS score and resulting in noticeable signal degradation (Figure \ref{fig:das_visual2}d). The SwinV2-adapted model (LS = 0.201) also exhibits limitations on this sample, failing to effectively suppress irregular and horizontal coherent noise (Figure \ref{fig:das_visual2}g).

  Examination of the removed noise sections in Figure~\ref{fig:das_visual1} further corroborates these observations. While all methods exhibit varying degrees of signal leakage due to the highly contaminated nature of DAS recordings, ABDIP preserves the most effective signal on this specific sample. However, considering the average quantitative metrics across the entire dataset, where DAS-N2N and our DINOv3-adapted model outperform ABDIP, demonstrates that DINOv3 provides a stable and robust general prior for the application of the VFMs in the DAP-VSP denoising task. It achieves a balance between signal preservation and background noise suppression across varying conditions, rivaling dedicated weakly-supervised or unsupervised DAS-VSP denoising methods. Consequently, given its strong universal structural priors, exceptional zero-shot adaptability, and competitive overall quantitative performance, we have selected DINOv3 as the default encoder for our proposed framework fine-tuning in all subsequent experiments.

  \subsection{Generalization on Unseen Field Data}

    To further evaluate the generalization capability of the proposed framework, we conducted comparative experiments on two unseen real-world datasets representing different geological environments: a land seismic line \cite[]{liu2013noncausal} and DAS-VSP data \cite[]{martuganova20223d}. These datasets exhibit noise characteristics and amplitude distributions that differ from the training data, providing a test of cross-domain robustness.
    
    The land seismic data consists of a 2D post-stack line extracted from a seismic survey in China, characterized by severe noise contamination \cite[]{birnie2021potential}. Different from the DAS-VSP sample in Figure~\ref{fig:das_visual1}a, the second DAS-VSP example extracted from the geothermal survey. It is distinguished by strong zigzag noise and random noise. We apply our proposed framework (DINOv3 PEFT+TTA, initially trained on the previous synthetic active seismic image dataset), the SFM Fine-tune \citep[trained on the same source dataset using the method of][]{sheng2025seismic}, and N2V (a self-supervised method trained directly on this specific field data) to the unseen data.

  \begin{figure}[htbp]
    \centering
    \includegraphics[width=1.0\textwidth]{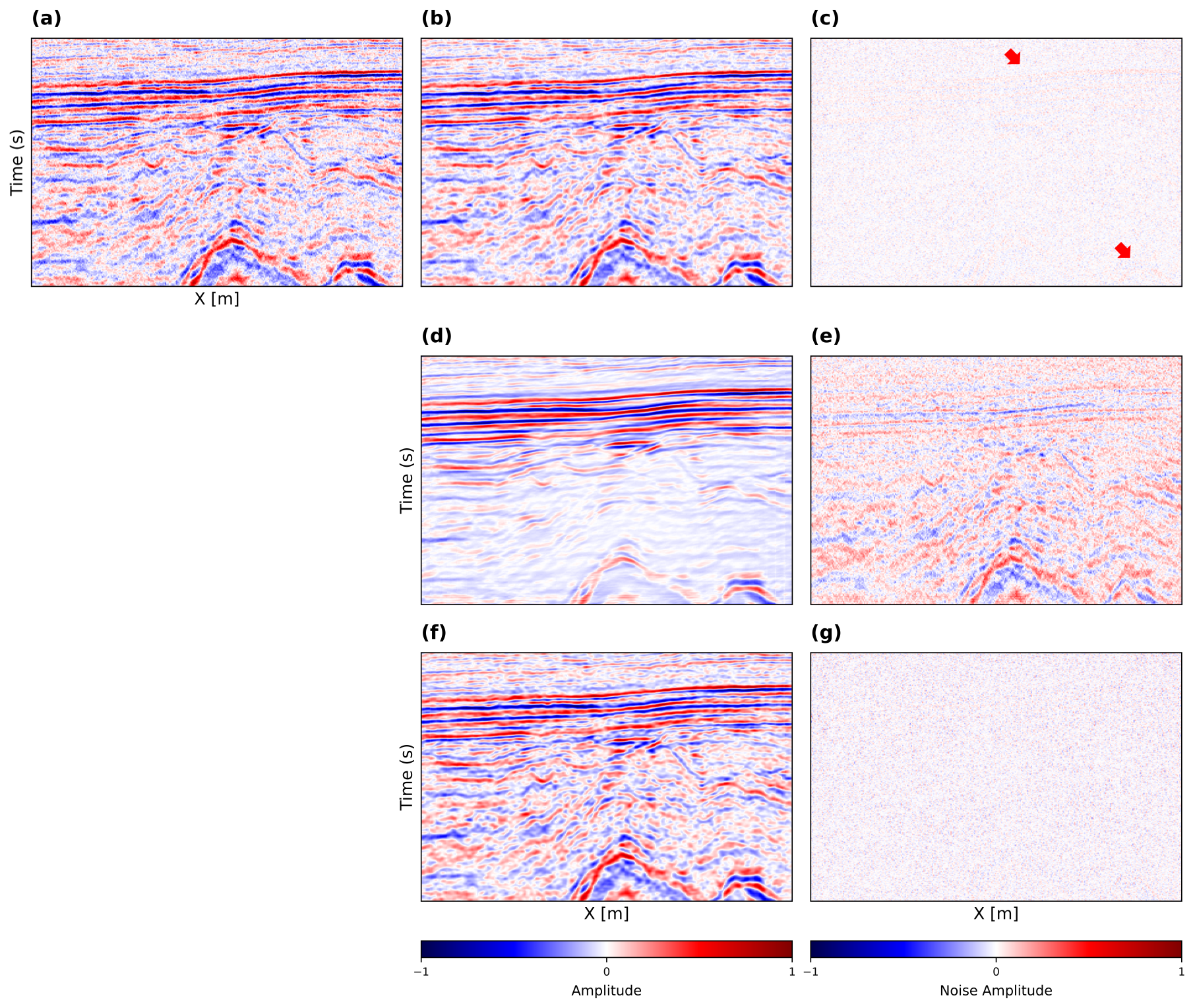}
    \caption{Comparison of denoising generalization on unseen land seismic data from China. (a) Raw data. Denoised results and corresponding removed noise for: (b-c) N2V (MS-SSIM = 0.9909, MS-SSIM-R = 0.5702), where red arrows indicate minor coherent signal leakage; (d-e) SFM Fine-tune (MS-SSIM = 0.6853, MS-SSIM-R = 0.6403), which shows serious signal leakage; and (f-g) our proposed framework (MS-SSIM = 0.9895, MS-SSIM-R = 0.5480). Note that our proposed framework achieves the best noise separation (lowest MS-SSIM-R) despite being applied to unseen data.}
    \label{fig:land_seismic_results}
  \end{figure}

  The denoising results are presented in Figure~\ref{fig:land_seismic_results}. Visual inspection reveals that the SFM Fine-tune suffers from severe domain shift when applied to unseen real-world data without adaptation. It exhibits over-smoothing in certain regions and severe signal leakage, with massive residual noise contamination (Figure~\ref{fig:land_seismic_results}e), resulting in poor quantitative metrics (MS-SSIM = 0.6853, MS-SSIM-R = 0.6403). In contrast, the self-supervised N2V method, which benefits from training directly on the target field data, effectively suppresses noise and achieves an excellent MS-SSIM of 0.9909. However, an analysis of its noise residual  reveals localized coherent signal leakage, highlighted by the red arrows, which translates to a higher MS-SSIM-R of 0.5702 (see Figure~\ref{fig:land_seismic_results}c). Remarkably, our proposed framework bridges the domain gap. Although its base weights were trained on a different synthetic dataset, the TTA mechanism enables it to dynamically adjust to the new data distribution. It effectively preserves intricate structural details and the relative amplitudes of different wave packets, achieving an MS-SSIM of 0.9895, which is competitive with the target-trained N2V. More importantly, it achieves the lowest MS-SSIM-R of 0.5480, demonstrating the cleanest noise separation  with minimal signal leakage compared to the competing methods (see Figure~\ref{fig:land_seismic_results}g).

\begin{figure}[htbp]
    \centering
    \includegraphics[width=0.85\textwidth,height=0.72\textheight,keepaspectratio]{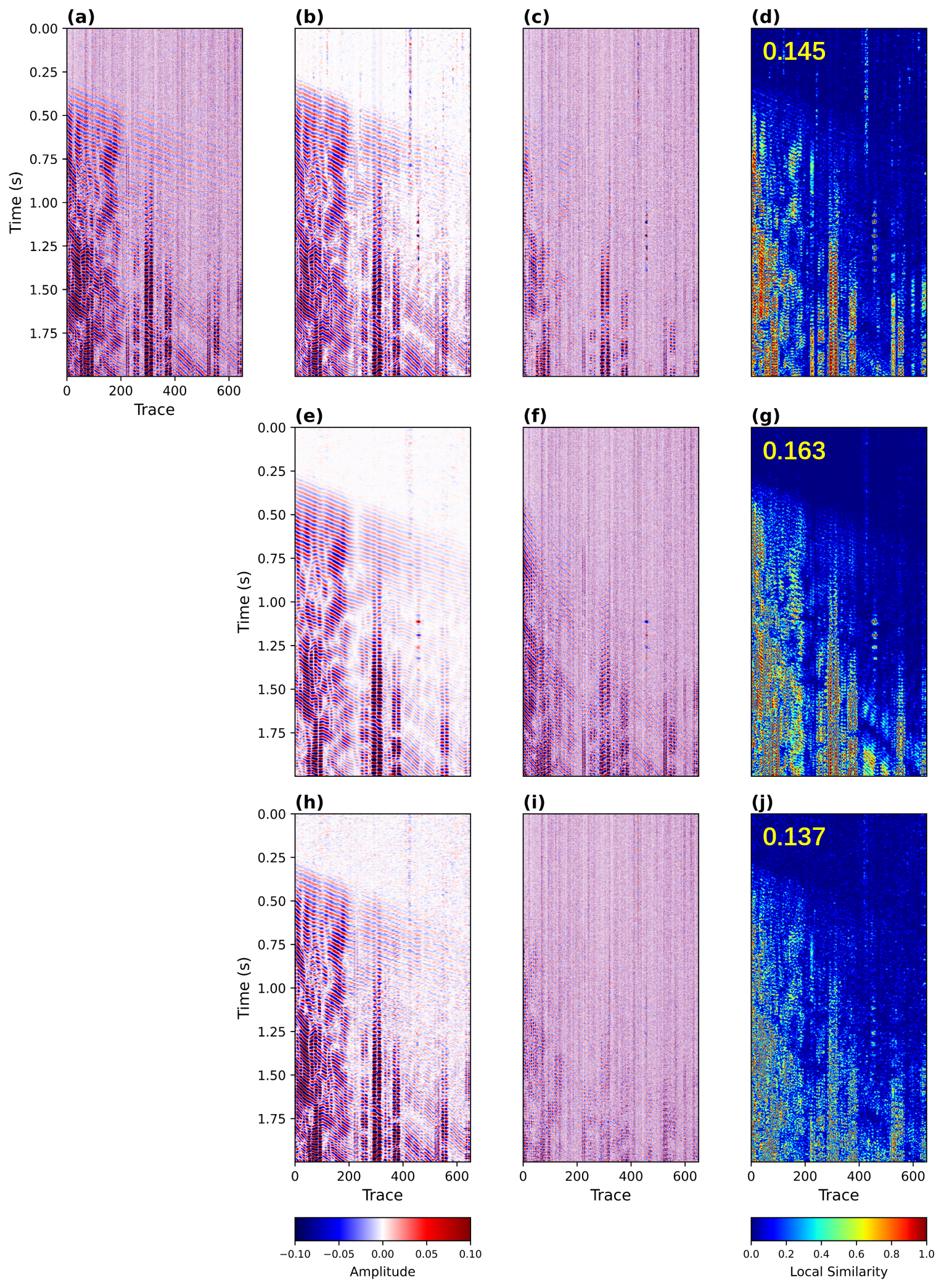}
    \caption{Generalization test on unseen DAS-VSP data from the Groß Schönebeck site. (a) Raw data. Denoised results, extracted noise residuals, and LS maps for: (b-d) DAS-N2N (LS = 0.145); (e-g) ABDIP (LS = 0.163); and (h-j) our proposed framework (LS = 0.137). Our framework achieves less signal leakage and clearer signal separation on this challenging sample.}
    \label{fig:das_vsp_results}
  \end{figure}

  The unseen DAS-VSP data is derived from a geothermal exploration at the Groß Schönebeck site in Germany. Compared to the Utah FORGE dataset, this data exhibits a significant domain shift, characterized by intense saw-tooth coupling noise, random noise, and amplitude variations of approximately three orders of magnitude. We apply our proposed framework (DINOv3 PEFT+TTA, initially fine-tuned on the Utah FORGE dataset), the specialized DAS-N2N, and the unsupervised ABDIP method to this unseen data to evaluate cross-domain generalization. The visual and quantitative results are shown in Figure~\ref{fig:das_vsp_results}.

    All three methods effectively suppress the severe random and saw-tooth coupling noise to a large extent. However, a closer inspection of the noise residual profiles and LS maps reveals differences in signal preservation under the unnseen data. The unsupervised ABDIP method struggles with the complex amplitude dynamics, exhibiting the highest signal leakage (LS = 0.163) and leaving noticeable coherent noise residuals. The weakly-supervised DAS-N2N performs better, yielding a lower LS of 0.145, but still demonstrates signal degradation in its LS map (Figure~\ref{fig:das_vsp_results}d). On the other hand, our proposed framework achieves the best signal preservation, evidenced by the lower values in its LS map (LS = 0.137). Although minor vertical noise traces remain in our denoised result (Figure~\ref{fig:das_vsp_results}h), our proposed framework outperforms both ABDIP and DAS-N2N in suppressing coherent noise and preserving true seismic amplitudes with less signal leakage into the noise profile (Figure~\ref{fig:das_vsp_results}i), which underscores the robustness and generalization of the denoising capability of our adapted VFM across diverse DAS-VSP geological settings.

  \subsection{Impact of Kurtosis-Guided TTA and Hyperparameter Analysis}

  To evaluate the contribution of the kurtosis-guided unsupervised TTA strategy, we conducted an ablation study comparing the denoising performance of the proposed framework under two conditions: with and without the TTA module. The evaluation was performed on two previously unseen field datasets, including active seismic image data and passive raw DAS-VSP data. In the case without TTA, during inference the model operates with fixed weights derived from the LoRA fine-tuning stage. When TTA is enabled, an additional unsupervised learning step is performed to iteratively update the LoRA parameters for input data patches, enabling adaptation to the target data distribution. Furthermore, we investigate the sensitivity of the TTA module to two key hyperparameters: the shift step, which controls patch overlap during inference, and the number of TTA epochs. Aside from these two variables, all other hyperparameters remain consistent with those detailed in Tables~\ref{tab:parameters} and \ref{tab:hyperparameters}.

  \begin{table}[htbp]
    \centering
    \caption{Ablation study results evaluating the impact of the kurtosis-based TTA strategy and key hyperparameters, including the shift step and TTA epochs. The table compares the quantitative denoising performance on unseen land seismic data using MS-SSIM and MS-SSIM-R, and on unseen DAS-VSP data using LS, between the inference phase without TTA and multiple configurations with TTA enabled. Note that $\uparrow$ ($\downarrow$) indicates that a larger (smaller) value represents better performance, with the best results shown in bold. Hyphens (-) denote omitted tests, as the denoising performance trends are clearly established.} 
    \label{tab:ablation_tta}
    \renewcommand*{\arraystretch}{1.1}
    \begin{tabular}{|lcccc|} 
      \hline
      \textbf{Method} & \textbf{Settings} & \textbf{MS-SSIM $\uparrow$} & \textbf{MS-SSIM-R $\downarrow$} & \textbf{LS $\downarrow$} \\
       & (Shift Step / Epochs) &  &  &  \\
      \hline
      w/o TTA & None & 0.7566 & 0.7491 & 0.319 \\
      \hline
      \multirow{11}{*}{w/ TTA} & $221 \times 221$ / $50$ & 0.8887 & 0.6307 & 0.139 \\
            & $112 \times 112$ / $50$ & 0.9190 & 0.6060 & 0.125 \\
            & $56 \times 56$ / $50$ & 0.9533 & 0.5790 & 0.101 \\
            & $28 \times 28$ / $50$ & 0.9762 & 0.5569 & 0.076 \\
            & $14 \times 14$ / $50$ & 0.9837 & 0.5654 & \textbf{0.059} \\
            & $7 \times 7$ / $50$ & \textbf{0.9892} & \textbf{0.5496} & 0.149 \\
            & $7 \times 7$ / $75$ & 0.9483 & 0.5563 & - \\
            & $7 \times 7$ / $25$ & 0.9833 & 0.5575 & - \\
            & $7 \times 7$ / $5$ & 0.9762 & 0.5570 & - \\
            & $14 \times 14$ / $25$ & - & - & 0.069 \\
            & $14 \times 14$ / $5$ & - & - & 0.095 \\
      \hline
    \end{tabular}
  \end{table}

    As shown in Table~\ref{tab:ablation_tta}, directly applying the fine-tuned model to unseen data without TTA results in limited performance (MS-SSIM of 0.7566 and LS of 0.319), which highlights the presence of domain shift in field datasets. Incorporating the TTA leads to consistent improvements across all quantitative metrics. The kurtosis-guided objective function effectively enables the model to differentiate non-Gaussian seismic signals from Gaussian random noise during inference, thereby recovering structural features that were previously distorted. 
    
    The ablation results also reveal insights into the influence of key hyperparameters. A smaller shift step provides denser overlapping patches, enabling finer local adaptation. For the land seismic dataset, performance improves as the step size decreases, reaching its best level at $7 \times 7$. In contrast, for the DAS-VSP dataset, the optimal resolution is $14 \times 14$ (LS = 0.059); further reduction to $7 \times 7$ leads to degraded performance (LS = 0.149). This discrepancy is likely associated with differences in spatial coherence between DAS-VSP noise and signals, where excessively small windows may cause the model to lose the broader contextual view of continuous optical noise or extended wavefields, leading to localized overfitting. Regarding the number of TTA epochs, we observe that performance generally increases up to 50 epochs. Notably, extending the training to 75 epochs for the $7 \times 7$ setting caused a performance drop (MS-SSIM decreased to 0.9483), indicating that excessive unsupervised updates may overwrite the valuable structural priors originally learned by the VFM.
    
    This observed performance gains are accompanied by increased computational cost. Smaller shift steps and larger epoch counts exponentially increase the number of inference iterations required per sample. Based on these observations, we recommend the following optimal configurations for practical applications: for active seismic image data, a $7 \times 7$ shift step with 50 epochs provides the absolute highest fidelity. For passive raw DAS-VSP data, a $14 \times 14$ shift step with 50 epochs yields optimal signal preservation. However, in scenarios where latency is a concern, the number of epochs may be reduced to 25 or even 5 (e.g., $7 \times 7$ / 5 epochs achieves an MS-SSIM of 0.9762) to cut processing time while still maintaining a competitive denoising that outperforms the non-TTA baseline.

  \subsection{Computational Efficiency Analysis}

  Computational efficiency is a critical consideration for practical deployment in seismic processing workflows. During the training phase, the proposed framework employs LoRA to update only 1.87 million parameters, whereas the SFM Fine-tune requires optimization of approximately 300 million parameters. This corresponds to a $99.39\%$ reduction in trainable parameters, significantly decreasing both memory consumption and training time.

  \begin{table}[htbp]
    \centering
    \caption{Computational efficiency comparison between the SFM Fine-tune configurations and our proposed framework (based on SFM and DINOv3 encoder). The table details the number of trainable parameters (T.Params), the total Fine-Tuning (FT) time, and the inference throughput (patches per second) measured on a single NVIDIA Tesla V100 GPU.}\label{tab:efficiency}
    \begin{tabular}{|lccc|} 
      \hline
      \multirow{2}{*}{\textbf{Method}} & \textbf{T.Params} & \textbf{Total FT Time} & \textbf{Infer. Throughput} \\
      \textbf{} & \textbf{(M)} & \textbf{(min.)} & \textbf{(patches per sec.)} \\
      \hline
      SFM Fine-tune & 306.38 & 112 & 7.61 \\
      SFM PEFT+TTA & 5.96 & 62 & 0.082 \\
      DINOv3 PEFT & \textbf{1.87} & \textbf{8} & \textbf{19.05} \\
      DINOv3 PEFT+TTA & \textbf{1.87} & \textbf{8} & 0.38 \\\hline
    \end{tabular}
  \end{table}

    As detailed in Table~\ref{tab:efficiency}, our proposed framework achieves a total fine-tuning time of just 8 minutes on a single NVIDIA Tesla V100 GPU, making it 14 times faster than the SFM Fine-tune (112 minutes). During inference, the performance depends on the chosen adaptation strategy. Without TTA, our lightweight DINOv3 PEFT model achieves a high inference throughput of 19.05 patches per second, which is approximately 2.5 times faster than the SFM Fine-tune (7.61 patches per second). However, the introduction of TTA reduces the throughput to 0.38 patches per second. This reduction occurs because TTA requires multiple unsupervised learning iterative updates for specific patches during inference. Nevertheless, this is still more efficient than applying PEFT+TTA to the SFM backbone, which drops to 0.082 patches per second.

    Therefore, in actual deployments for in-domain seismic data that shares a similar distribution with the training set, we utilize the PEFT configuration as shown in Table~\ref{tab:parameters} (without TTA), which guarantees optimal processing speed (19.05 patches/s) while maintaining denoising fidelity. Conversely, for out-of-domain or completely unseen field data (e.g., cross-site surveys) where structural preservation is the priority, we deploy the PEFT+TTA configuration. Crucially, the base PEFT model only needs to be trained once on an available dataset. The subsequent TTA process automatically bridges the domain gap during inference without requiring manual labeling, new paired data collection, or offline retraining, thereby offsetting the slower per-patch inference speed with savings in human labor and data preparation.

  However, compared to unsupervised learning methods such as ABDIP, our approach does not possess a speed advantage during the training and inference phase. For instance, ABDIP performs denoising on unseen DAS-VSP data by conducting unsupervised training directly on the target data, and its concise network architecture yields fast processing times. However, our proposed framework demonstrates generalization capability and few-shot robustness. By fine-tuning just once on a limited synthetic or real dataset, it can effectively denoise complex real-world recordings from diverse geological surveys, achieving signal separation that rivals or exceeds dedicated advanced algorithms. Additionally, its universal nature allows for adaptation across seismic data modalities, ranging from active land surveys to passive DAS-VSP arrays, which provides greater flexibility and lays a scalable foundation for a broader range of downstream geophysical tasks.

  \section{Discussion}

  The experimental results demonstrate that VFMs, when adapted, provide strong representation capabilities for seismic denoising across multiple acquisition types. While the proposed framework achieves robust generalization, several limitations of the current methodology remain, which warrant further investigation. First, the scope of this study is currently limited to denoising, thereby excluding other critical seismic processing tasks such as interpolation and demultiple. Second, the reliance on fixed $224 \times 224$ input patches constitutes a constraint; although the employed overlap-stitch method mitigates boundary discontinuities, processing large volumes via patching can still introduce edge artifacts. Furthermore, the methodology processes 2D sections, which implies that the 3D spatial continuity of volumetric seismic data is not fully exploited. Finally, regarding computational latency, the current TTA implementation requires a multi-epoch adaptation phase before inference. This adaptation period presents a bottleneck for real-time monitoring applications, such as microseismic event detection, where immediate feedback is essential.

  To address these challenges and broaden the practical applicability of the proposed framework, future research should focus on exploring more flexible architectural designs. Extending the current 2D patch-based adaptation to fully exploit 3D spatial continuity, as well as optimizing the adaptation phase to reduce computational latency, are crucial steps for real-time field deployments. Additionally, investigating the scalability of these adapted VFMs to encompass a broader range of seismic processing workflows beyond denoising remains a valuable and open direction for the geophysical research.

  \section{Conclusions}

  This study presents an efficient framework that bridges computer vision and exploration geophysics by effectively repurposing pre-trained VFMs for seismic data denoising. By integrating the DINOv3 encoder with PEFT via LoRA, we demonstrate that general-purpose VFMs can achieve strong performance in both active seismic and passive DAS-VSP denoising tasks. Specifically, LoRA proves superior to full fine-tuning as a cost-effective adaptation strategy; it effectively preserves the pre-trained visual features while reducing computational overhead, a capability that is important for label-scarce geophysical scenarios. Furthermore, our findings highlight that DINOv3 outperforms architectures like SwinV2 and domain-specific models such as the SFM. Its balanced global-local feature extraction renders it ideal for capturing both large-scale structures and fine-grained wavefields in conventional seismic data and DAS-VSP data within a single framework.

  Another contribution of this work is the Kurtosis-Guided TTA module, which transforms the denoising model from a static inference engine into a dynamic, self-calibrating system. By leveraging the high kurtosis of sparse seismic signals, the patch selection strategy identifies information-rich regions to adapt to unseen noise distributions without requiring manual labels. This capability mitigates domain shifts and alleviates the long-standing challenge of domain generalization in geophysical deep learning.

  Ultimately, our findings suggest that the future of seismic processing may not lie in training ever-larger domain-specific models from scratch, but in the intelligent adaptation of foundation models. Our proposed framework provides a cost-effective, scalable, and robust solution for data-intensive challenges in exploration seismology, paving the way for the broader deployment of foundation models in subsurface imaging and monitoring.

  \section{Acknowledgements}

  We extend our sincere gratitude to the Delphi Consortium at Delft University of Technology and King Fahd University of Petroleum \& Minerals (KFUPM) for their support and collaboration, we also thank KFUPM for facilitating the two-month research visit that contributed to this work. This work is supported by computing time awarded on the Cyclone supercomputer of the High-Performance Computing Facility, CyI. 

  Regarding data, code, and technical support, we are highly grateful to Dr. Yangkang Chen for making the processed DAS-VSP dataset and corresponding processing codes open-source. We further extend our appreciation to Mr. Hanlin Sheng for providing the open-source SFM code and fine-tuning datasets, Dr. Claire Birnie for sharing the open-source N2V denoising code and test field data, and Dr. Sacha Lapins for making the DAS-N2N code publicly available. Finally, we sincerely thank all the individuals and organizations who generously release their pre-trained model weights and open-source codebases to the community.

  This work was funded in part by the European Union's Horizon 2020 research and innovation programme under Grant Agreement No. 101034267. Additional support was provided by internal funding from the Department of Geosciences at KFUPM.

  \section{Data and Materials Availability}
  The datasets and processing code supporting the findings of this study are available as follows:
  \begin{itemize}
    \item \textbf{Active seismic image data:} the open-source synthetic and field active seismic denoising datasets used in this study are provided by \cite{sheng2025seismic} and can be accessed via their associated repository: \url{https://github.com/shenghanlin/SeismicFoundationModel?tab=readme-ov-file#rocket-model-zoo--data-release}; and the 2D land seismic line for generalization testing is provided by \cite{birnie2021potential}, which can be found in \url{https://drive.google.com/file/d/1Ln7Vx2whs_zSlucLEDIEQH_XgY2FBX-P/view?usp=drive_link}.
    \item \textbf{Passive raw DAS-VSP data:} the passive raw DAS-VSP data is derived from the public Utah FORGE field site (\url{https://utahforge.com/project-data-dashboard/}), a collected dataset containing seismic records comes from \url{https://github.com/chenyk1990/dasdenoising-dataonly/tree/main/mat_raw}, and ground truth labels were generated using workflows described in \cite{dasdenoising2023}, which can be accessed via repository: \url{https://github.com/TCCS-CODES/dasdenoising}. Data (used in our study) from the Groß Schönebeck geothermal site can be downloaded directly at \url{https://github.com/cuiyang512/Unsupervised-DAS-Denoising/blob/main/data/slice_german_1.npy}.
    \item \textbf{Pre-trained models:} the pre-trained SFM, SwinV2 and DINOv3 weights are available in their respective public repositories: \url{https://github.com/shenghanlin/SeismicFoundationModel?tab=readme-ov-file#rocket-model-zoo--data-release} for SFM-Large ($224 \times 224$); \url{https://github.com/microsoft/Swin-Transformer?tab=readme-ov-file#main-results-on-imagenet-with-pretrained-models} for SwinV2-B ($224 \times 224$, pretrained on ImageNet-22K); \url{https://huggingface.co/facebook/dinov3-vits16-pretrain-lvd1689m} for DINOv3 ViT-S/16.
    \item \textbf{Related code:} the proposed framework is implemented using PyTorch: \url{https://pytorch.org/} and the Hugging Face PEFT library: \url{https://github.com/huggingface/peft}. The training and inference code for the methods used in comparison can be found in the following repositories, SFM: \url{https://github.com/shenghanlin/SeismicFoundationModel}; N2V: \url{https://github.com/cebirnie92/SelfSupervisedDenoising_Tutorials}; DAS-NSN: \url{https://github.com/sachalapins/DAS-N2N-torch}; ABDIP: \url{https://github.com/cuiyang512/Unsupervised-DAS-Denoising}.
  \end{itemize}

  \bibliographystyle{seg} 
  \bibliography{references}

  \end{document}